\documentclass[aps,prb,twocolumn]{revtex4}
\usepackage{graphicx}
\usepackage{amssymb}
\begin{document}
\bibliographystyle{apsrev}

\title{Evaporation of Lennard-Jones Fluids}
\author{Shengfeng Cheng}
\email{sncheng@sandia.gov}
\affiliation{Sandia National Laboratories, Albuquerque, NM 87185, USA}
\author{Jeremy B. Lechman}
\affiliation{Sandia National Laboratories, Albuquerque, NM 87185, USA}
\author{Steven J. Plimpton}
\affiliation{Sandia National Laboratories, Albuquerque, NM 87185, USA}
\author{Gary S. Grest}
\affiliation{Sandia National Laboratories, Albuquerque, NM 87185, USA}

\date{\today}

\begin{abstract}
Evaporation and condensation at a liquid/vapor interface 
are ubiquitous interphase mass and energy transfer phenomena that 
are still not well understood.
We have carried out large scale molecular dynamics simulations of 
Lennard-Jones (LJ) fluids composed of monomers, dimers, 
or trimers to investigate these processes with molecular detail.
For LJ monomers in contact with a vacuum, 
the evaporation rate is found to be very high 
with significant evaporative cooling and 
an accompanying density gradient in the liquid domain 
near the liquid/vapor interface. 
Increasing the chain length to just dimers 
significantly reduces the evaporation rate. 
We confirm that mechanical equilibrium plays a key role
in determining the evaporation rate
and the density and temperature profiles
across the liquid/vapor interface.
The velocity distributions of evaporated molecules 
and the evaporation and condensation coefficients 
are measured and compared to the predictions of an existing model 
based on kinetic theory of gases. 
Our results indicate that for both monatomic and polyatomic molecules, 
the evaporation and condensation coefficients are equal when 
systems are not far from equilibrium and smaller than one, 
and decrease with increasing temperature.
For the same reduced temperature $T/T_c$, 
where $T_c$ is the critical temperature,
these two coefficients are higher for 
LJ dimers and trimers than for monomers, in contrast to 
the traditional viewpoint 
that they are close to unity for monatomic molecules and 
decrease for polyatomic molecules. 
Furthermore, data for the two coefficients collapse onto a master curve when
plotted against a translational length ratio between the liquid and
vapor phase.
\end{abstract}


\maketitle

\noindent{\bf I. INTRODUCTION}
\bigskip

The inverse processes of evaporation and condensation are 
of fundamental importance in natural phenomena and engineering applications. 
In both processes, heat and mass transfer between liquid and vapor phases.\cite{schrage53}
The key physical quantities to determine are the interphase mass 
and energy transfer rates. 
There have been a number of theoretical analyses, using either kinetic theory
\cite{schrage53,pao71a,pao71b,siewert73,sone73,sone78b,
cipolla74,labuntsov79,koffman84,aoki90}
or nonequilibrium thermodynamics,\cite{bedeaux86,bedeaux90,bedeaux99}
to predict these rates as well as the density, temperature, 
and pressure profiles in the vapor phase.
In the framework of kinetic theory of gases, 
the problem was typically formulated based on 
the Boltzmann-Bhatnagar-Gross-Krook-Welander (BBGKW) equation for the vapor, 
with the liquid surface temperature and 
the temperature, pressure, and velocity of vapor far away from
the interface as boundary conditions. 
To obtain solutions of the BBGKW equation, 
one further assumption of the liquid/vapor interface was usually made, 
i.e., all molecules approaching the interface 
completely condense into the liquid phase, while
molecules evaporated from the liquid surface 
have a Maxwell-Boltzmann (MB) distribution corresponding to the
saturated vapor at the liquid temperature. 
This is equivalent to assuming that the evaporation and 
condensation coefficients are unity
since the evaporation (condensation) coefficient is defined as 
the ratio of an experimental evaporation (condensation) rate
to a theoretical maximum rate given by the Hertz-Knudsen (HK) equation, 
which is exactly the rate corresponding to a MB distribution.
Though the evaporation and condensation coefficients are closely related 
and have the same value for an interface in equilibrium,
they could be substantially smaller than unity 
and have different values for nonequilibrium interfaces. 
The above assumption of the boundary condition at 
the liquid/vapor interface might not
be realistic and recently some theoretical work emerged attempting
to replace it with more physical ones.\cite{bond04,frezzotti05,holyst08,holyst09,caputa11}

The kinetic theory of
evaporation and condensation was challenged 
in a recent experiment of Fang and Ward.\cite{fang99a} 
They measured the temperature profile to within
one mean free path ($\sim 19\mu m$) of the interface of an
evaporating liquid and found a discontinuity in temperature across
the interface that was much larger in magnitude and in the opposite direction
to that predicted by the kinetic theory or nonequilibrium thermodynamics.
This disagreement led Fang and Ward to conclude that 
the boundary conditions traditionally assumed for 
the BBGKW equation were unphysical and they developed
a statistical rate theory of evaporation flux to explain their
experimental observations.\cite{fang99b,fang99c}

To elucidate the physics of liquid/vapor interface during
evaporation and condensation, 
more detailed measurements are still needed 
at even smaller scales and with more refined resolutions.
However, the liquid/vapor interface is difficult to probe experimentally 
and available data can only provide limited information of 
the microscopic detail of an evaporating interface.
In the past several decades, although many measurements have been reported on 
the evaporation and condensation coefficients of various materials, 
it still remains unclear how these results can be related to the 
molecular processes occurring at the liquid/vapor interface. 
Furthermore, reported values are often scattered in such a large range 
even for the same material which
further complicates their theoretical interpretation.
For example, the reported measurements of the evaporation coefficient of water
range from $0.01$ to $1.0$,\cite{eames97,marek01}
which exemplifies the difficulty to obtain accurate values of these coefficients, 
let alone the molecular mechanism of evaporation and condensation. 

The development of computer simulation techniques, 
particularly the molecular dynamics (MD) method, has
enabled a number of studies of the evaporation and condensation processes at the molecular scale,
\cite{matsumoto88,matsumoto89,matsumoto92,yasuoka94,matsumoto94,
tsuruta95,tsuruta99,tsuruta03,tsuruta05,
anisimov97,anisimov99,rosjorde00,rosjorde01,meland04,fujikawa04,holyst08,holyst09}
which have advanced substantially our understanding of 
these interphase mass and energy transfer phenomena. 
Matsumoto {\it et al.} studied the liquid-vapor interfaces 
of argon, water, and methanol, and obtained 
the evaporation and condensation coefficients 
that agreed reasonably well with experimental values.\cite{matsumoto88,matsumoto89,matsumoto92,yasuoka94,matsumoto94} 
Tsuruta {\it et al.} studied the condensation coefficient 
of Lennard-Jones (LJ) fluids by injecting test molecules to 
bombard the liquid/vapor interface 
and measured the reflecting probability, from which 
the condensation coefficient was deduced.
\cite{tsuruta95,tsuruta99,tsuruta03,tsuruta05} 
Their results confirmed that the evaporation and condensation coefficients 
are equal in equilibrium systems, 
but revealed that the condensation probability generally depends on 
the normal component of the kinetic energy of incident molecules. 
This is in contrast to the common assumption that 
the condensation probability is constant for 
vapor molecules that hit the liquid surface.
They also measured the velocity distributions of 
the evaporated and reflected molecules 
at the liquid/vapor interface. 
Inspired by the MD results, Tsuruta {\it et al.} developed an expression for 
the velocity-dependent condensation probability 
and later justified this expression using the transition state theory,
\cite{tsuruta99, tsuruta03}
which was used earlier to estimate 
the condensation coefficient by Fujikawa and Maerefat.\cite{fujikawa90} 
Anisimov {\it et al.} also used MD to investigate 
the evaporation of LJ fluids and the properties of liquid/vapor interface, 
and extended their studies to the case of high-rate evaporation.\cite{anisimov97,anisimov99}

Recently, Rosjorde {\it et al.} used MD simulations to study 
LJ/spline fluids and found evidence for the hypothesis of 
local equilibrium at a liquid/vapor interface.\cite{rosjorde00, rosjorde01}
They also measured transfer coefficients of the mass and energy fluxes 
and found that they agreed with kinetic theory from the triple point
to about halfway to the critical point. 
They suspected that
the disagreement between kinetic theory predictions 
and experimental results of Fang and Ward 
was due to the fact that the theory dealt with monatomic fluids, 
while in the experiment polyatomic fluids were used.
Meland {\it et al.} also studied LJ/spline fluids and compared
their MD results with gas-kinetic calculations.\cite{meland04}
They found that the evaporation and condensation coefficients are 
not equal outside equilibrium and there is a significant 
drift velocity in the distribution function at the interphase 
for both net evaporation and net condensation.
Ishiyama {\it et al.} used MD simulations of LJ fluids to check the validity of 
kinetic boundary condition for the BBGKW equation and found 
that the condensation coefficient is close to unity below the triple-point
temperature and decreases gradually as the temperature rises.\cite{fujikawa04}

More recently, Holyst and Litniewski studied the evaporation of nanodroplets
and demonstrated that the evaporation process is limited
by the heat transfer and energy balance condition.\cite{holyst08}
This finding challenges the basic assumption of kinetic theory 
that the evaporation flux is determined by the diffusion of mass in the vapor phase.
In another study, they simulated a LJ liquid film evaporating into
a vacuum and measured the density, temperature, and pressure profiles.\cite{holyst09} 
Holyst and Litniewski found that mechanical equilibrium is established very quickly
and derived an expression for the mass flux that 
described their simulation results much better 
than the frequently used HK formula.

In these previous studies, only simple LJ fluids composed of monomers were used. 
However, it is well known that the vapor pressure of 
a LJ fluid is much higher than those of real liquids. 
Though a high vapor pressure implies that LJ liquids are relatively easy to evaporate, 
it does not necessarily mean that the evaporation coefficient 
is close to unity because the corresponding theoretical maximum flux 
is also large. 
In fact, some previous simulations found that at a moderate temperature the evaporation and 
condensation coefficients for simple LJ liquids are around $0.8$.
\cite{yasuoka94,tsuruta99,rosjorde01}

In this paper, we first simulate the evaporation of LJ monomers and then
extend MD simulations to the evaporation of LJ fluids 
composed of dimers and trimers. This allows us to examine the effect of 
molecular composition on the evaporation rate and 
the evaporation and condensation coefficients.
The paper is organized as follows. In Sec.~II, 
the simulation methodology and the procedure for creating the
liquid/vapor interface and removing atoms to implement controlled evaporation are introduced.
Results on phase diagrams and surface tensions of dimers and trimers 
are presented in Sec.~III.
In Sec.~IV results on density and temperature profiles are presented.
The measured evaporation rates are found to agree well with 
the modified HK expression derived by Holyst and Litniewski.\cite{holyst09}
The role of mechanical equilibrium during evaporation is also examined.
Then in Sec.~V we measure the velocity distributions
of the evaporated and condensed molecules and 
compare them to predictions of a kinetic model of evaporation based on 
the transition state theory of Tsuruta {\it et al.}.
The evaporation and condensation coefficients are also determined.
A brief summary and conclusions are included in Sec.~VI.

\bigskip\noindent{\bf II. SIMULATION METHODOLOGY}
\bigskip

We carried out large scale MD simulations of three simple liquids. 
In all three cases the interaction between atoms is described by 
the standard LJ 12-6 interaction
\begin{equation}
U(r)=4\epsilon\left[ (\sigma/r)^{12}-(\sigma/r)^6-(\sigma/r_c)^{12}+(\sigma/r_c)^6 \right],
\label{LJPotential}
\end{equation}
where $r$ is the distance between two atoms, 
$\epsilon$ is the unit of energy, and $\sigma$ is the diameter of the atom. 
The interaction is truncated at $r_c=2.5\sigma$. 
While commonly used to simulate liquids, 
the LJ model has one serious drawback 
when modeling the evaporation process. 
Namely it has a very high vapor pressure. 
This results in a very large vapor density 
which does not correspond well to most liquids. 
One simple way to make the model more realistic 
but still retain the simplicity of the LJ interaction 
is to model the solvent as short chains of LJ atoms. 
In this paper we show that by modeling dimers that consist of two atoms 
and trimers consisting of a linear chain of three atoms, 
we obtain a more reasonable low vapor density.
In dimer and trimer molecules, the bonded atoms are connected by an additional 
finitely extensible nonlinear elastic (FENE) potential,
\begin{equation}
U_{\rm FENE}(r)= -\frac{1}{2}k R_{0}^{2} {\rm ln}
\left[ 1-(r/R_0)^2 \right],
\label{FENEPotential}
\end{equation}
where $k=30k_{\rm B}T/\sigma^2$, 
$T$ is the temperature, 
$k_{\rm B}$ is the Boltzmann constant, 
and $R_0=1.5\sigma$.\cite{kremer90}

\begin{figure}[htb]
\centering
\includegraphics[width=3.25in]{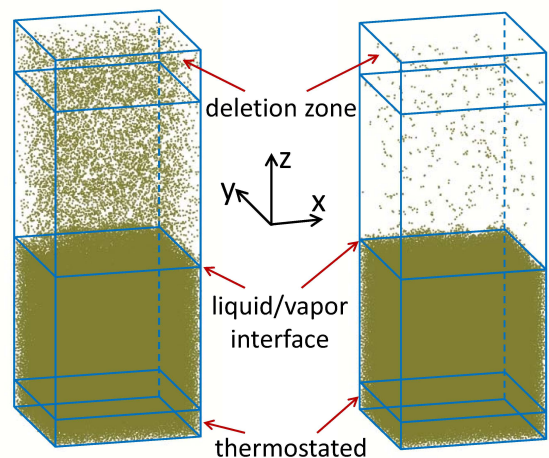}
\caption{(Color online) Liquid/vapor equilibrium at $T=0.9\epsilon/k_{\rm B}$ 
for LJ monomers (left) and dimers (right). The vapor density is clearly lower for dimers.}
\label{AtomicConfig}
\end{figure}

All MD simulations were performed using the LAMMPS simulation package.\cite{plimpton95}
Each system started with 
$N=3,000,000$ atoms in a parallelepiped of dimensions $L_x \times L_y \times L_z$, 
where $L_x=L_y=200\sigma$. 
The simulation cell was periodic in the $x$ and $y$ directions 
and had upper and lower confining walls in the $z$ direction as illustrated in Fig.~\ref{AtomicConfig}.
Both upper and lower $z$-walls interacted with the monomers with 
a LJ 9-3 potential, which depends only on the distance $z$ from the wall,
\begin{equation}
U(z)=\epsilon_{\rm w}\left[ \frac{2}{15}(\sigma/z)^{9}-(\sigma/z)^3 
- \frac{2}{15}(\sigma/z_c)^{9}+(\sigma/z_c)^3 \right],
\label{WallPotential}
\end{equation}
where $\epsilon_{\rm w} = 2\epsilon$. 
For the lower wall the interaction was truncated at $z_c=2.5\sigma$, 
while at the upper wall was purely repulsive with $z_c=0.72\sigma$. 
Each liquid film was constructed by placing $N$ atoms randomly in the simulation cell with
$L_z$ chosen so that the system was near its bulk liquid density. Overlaps were removed by running
a short simulation with the ``nve/limit'' fix in LAMMPS turned on.\cite{lammps}
This limits the
maximum displacement of an atom per time step and is a very efficient way to remove
overlaps between atoms. The system was then equilibrated at pressure
$P=0$ by adjusting the position of the upper wall.
After the equilibration, the  
position of the upper wall was increased by at least 
$50\sigma$ to form a liquid/vapor interface as shown in Fig.~\ref{AtomicConfig}.
The system was then allowed to come to equilibrium with its vapor 
before the evaporation process was initiated. 

The equations of motion were integrated using a velocity-Verlet algorithm 
with a time step $\delta t =0.005\tau$ for the dimers and trimers and $0.01\tau$ for the monomers,
where $\tau=\sigma(m/\epsilon)^{1/2}$ and $m$ is the monomer mass. 
However, in simulations to investigate the mechanical equilibrium during evaporation, 
the time step was reduced to $0.001\tau$ in order to get the 
small pressures accurate in the liquid phase.
During the equilibration, $T$ was held constant by a Langevin thermostat weakly 
coupled to all atoms with a damping constant $\Gamma=0.1\tau^{-1}$. 
Once the liquid/vapor interface was equilibrated, 
the Langevin thermostat was removed except
for those atoms within $15\sigma$ of the lower boundary at $z=0$. 
We refer to the liquid temperature in this region 
as the bulk temperature $T_b$.
The system was then equilibrated for  
an additional $20,000$ to $100,000$ time steps
before the evaporation process was initiated. 

To model evaporation, a deletion zone of thickness $20\sigma$ approximately $50\sigma$ 
above the starting liquid/vapor interface was defined where atoms were removed at a specified rate. 
For dimer and trimer systems, once one atom from a molecule was removed, 
the entire molecule was removed.
In this paper, the evaporation rate is defined as the number of atoms 
removed per unit time and area.
Therefore, one removed dimer (trimer) molecule contributes two (three) 
evaporated atoms.
In simulations with a controlled evaporation rate, 
$n$ atoms in the deletion zone were removed every $N_t$ time steps.
Typically, $n=10$ and $N_t=100$ or $1000$ for monomers and dimers. 
Since each trimer molecule contains $3$ atoms, setting
$n=9$ and $N_t=90$ or $900$ for trimer systems ensured the same evaporation rates as for the other
two systems.
By removing all atoms that entered the deletion zone, 
a system in contact with a vacuum was effectively modeled. 
Since only those atoms within $15\sigma$ of the lower
wall were coupled to the thermostat during the evaporation
process, the thermostat did not affect the evaporation process at the liquid/vapor interface. 
To avoid any finite size effects we only analyzed evaporation data 
for films thicknesses larger than $\sim 50\sigma$.

In our simulations, the density and temperature profiles in the simulation cell were measured. 
For liquid/vapor equilibrium cases, time-averages of the density and surface tension were carried out 
to determine the liquid/vapor phase diagrams.
For systems undergoing evaporation, instantaneous density and temperature distributions
were calculated. 
Since these systems are translationally invariant in the $x$-$y$ plane, 
both quantities only depend on $z$ and 
results presented here were averaged over $x$ and $y$ directions.

The meaning and definition of temperature in non-equilibrium systems 
still have many ambiguities.\cite{cacas03} 
In this paper, we measured a local temperature
$T(z)$ in two ways. In one, $T(z)$ is taken as the mean kinetic energy
of atoms locating in the spacial region from $z-\Delta z$ to $z+\Delta z$, 
where $\Delta z$ is typically $0.5\sigma$.
The corresponding expression is 
\begin{equation}
T(z)=\frac{m}{3N_{l}k_{\rm B}}\sum_{z-\Delta z}^{z+\Delta z}v^2,
\label{TempDef}
\end{equation}
where $N_l$ is the number of atoms in the region and
$v$ is the atomic velocity.
Another definition of the local temperature is based on 
velocities relative to the possible advective motion induced by evaporation
and can be written as 
\begin{equation}
T_r(z)=\frac{m}{3N_{l}k_{\rm B}}\sum_{z-\Delta z}^{z+\Delta z}(v-\overline{v})^2,
\label{TempDefRelative}
\end{equation}
where $\overline{v}$ is the mean atomic velocity in the spatial region 
from $z-\Delta z$ to $z+\Delta z$.
Clearly, the two definitions give identical results in equilibrium 
since $\overline{v}=0$ there.
However the results are generally different in nonequilibrium states.
This will be discussed further in Sec.~IV.

\bigskip\noindent{\bf III. LIQUID/VAPOR EQUILIBRIUM}
\bigskip

\begin{figure}[htb]
\centering
\includegraphics[width=2.5in]{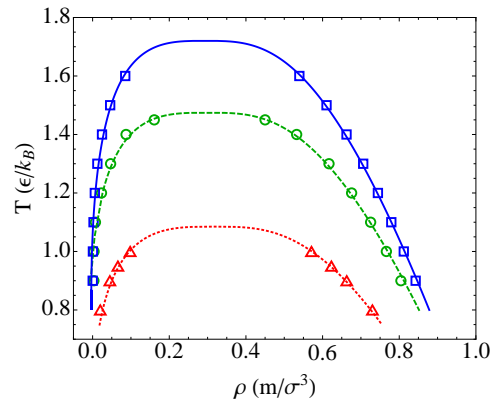}
\caption{(Color online ) Liquid/vapor phase diagram for LJ monomers (triangles), 
dimers (circles), and trimers (squares).}
\label{PhaseDiag}
\end{figure}

For the system composed of LJ monomers, the liquid/vapor phase diagram is well 
known.\cite{nijmeijer88,adams91,smit92,johnson93,wilding95,potoff98,sides99,shi01}
However, this is not the case for dimers and trimers. 
Therefore for these two systems we also simulated the liquid/vapor coexistence
by removing the walls in the $z$ direction and 
replacing them with periodic boundary conditions, 
which resulted in two liquid/vapor interfaces in the simulation cell. 
The liquid and vapor densities $\rho_L$ and $\rho_V$ 
respectively along the coexistence curve for 
all three systems are shown in Fig.~\ref{PhaseDiag}. 
The liquid/vapor critical temperature $T_c$ for each system is 
determined from fitting the measured densities to
\begin{equation}
\begin{array}{lll}
\rho_L + \rho_V & = & a-b~T, \\
\rho_L - \rho_V & = & A(1-T/T_c)^{\beta^\prime},
\end{array}
\label{PhaDiagFit}
\end{equation}
where $a$, $b$, and $A$ are fitting parameters.\cite{adams91}
The critical exponent $\beta^\prime = 0.318$ is for Ising systems that are in the
same universality class as simple fluids; thus it is fixed to this value in the fitting. 
The best fits using Eq.~(\ref{PhaDiagFit}) 
give the critical temperature 
$T_c = 1.085$, $1.475$, and $1.720\epsilon/k_{\rm B}$, and 
the critical density  
$\rho_c = 0.316$, $0.299$, and $0.294 m/\sigma^3$
for the monomer, dimer, and trimer systems, respectively.

\begin{figure}[htb]
\centering
\includegraphics[width=2.5in]{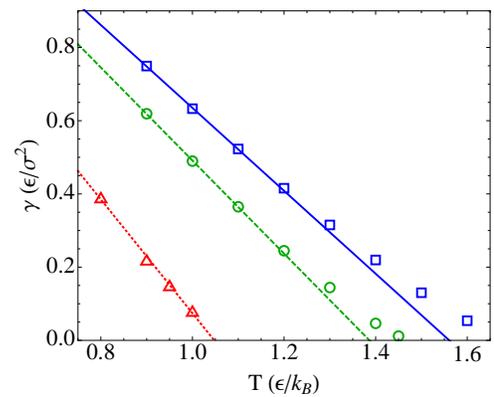}
\caption{(Color online) Surface tension $\gamma$ as a function of temperature 
for LJ monomers (triangles), 
dimers (circles) and trimers (squares). The results for monomers
are taken from Ref.~\onlinecite{sides99}.}
\label{SurfTen}
\end{figure}

The surface tension of the liquid/vapor interface was determined by 
measuring the stress tensor. Since the two interfaces are parallel to the
$x$-$y$ plane, the surface tension $\gamma$ is given by the
Kirkwood-Buff formula,\cite{kirkwood49}
\begin{equation}
\begin{array}{lll}
\gamma & =  &\frac{1}{2}\int_{0}^{L_z} \left[ p_{zz}(z)-(p_{xx}(z)+p_{yy}(z))/2\right] {\rm d}z \\
& = & \frac{L_z}{2} \left[ P_{z}-(P_{x}+P_{y})/2\right],
\end{array}
\label{KBEq}
\end{equation}
where $p_{xx}(z)$, $p_{yy}(z)$, and $p_{zz}(z)$ are 
the three diagonal components of the stress tensor and
$P_{x}$, $P_{y}$, and $P_{z}$ are the spatially averaged pressure in each direction.
The factor $1/2$ before the integral comes from the fact that there are two interfaces.
Far from the liquid/vapor interface, both the liquid and vapor phases are homogeneous 
and isotropic, and all diagonal components of the stress tensor are the same and equal to the
hydrostatic pressure. 
Therefore in these regions the integrand in Eq.~(\ref{KBEq}) is zero
and does not contribute to the integral.
Near the interface whose normal is along the $z$-direction, 
$p_{xx}(z)$ and $p_{yy}(z)$ are less than $p_{zz}(z)$, 
leading to a difference in $P_{x}$, $P_{y}$, and $P_{z}$.
The net outcome is a nonzero surface tension of the liquid/vapor interface. 
Results for $\gamma$ are shown 
in Fig.~\ref{SurfTen} for LJ monomers, dimers, and trimers. 
As expected, $\gamma$ drops as temperature is raised and approaches zero as $T_c$ is reached.
When the temperature $T$ is well below $T_c$, 
the surface tension $\gamma$ roughly decreases linearly with $T$.
However, when $T_c$ is approached the rate of reduction of $\gamma$ becomes smaller.
It is expected $\gamma \sim (T_c-T)^{\nu}$ with $\nu$ as a critical exponent.
Results in Fig.~\ref{SurfTen} are consistent with $\nu > 1$.\cite{potoff00}

Figure~\ref{SurfTen} also shows that the surface tension is
smaller for monomers and larger for dimers and trimers.
This implies that the cohesion between molecules becomes stronger 
as the chain length gets longer.
Accordingly, at the same temperature the vapor densities of dimers and trimers are lower than that
of monomers, as illustrated in Fig.~\ref{AtomicConfig}.
Note that dimers and trimers considered in this paper are 
very flexible objects. It is not surprising that longer chains 
lead to stronger attractive interactions between molecules 
because there are more interacting sites and 
the counteractive effect of rotational degrees of freedom is suppressed.
We will see later that the stronger cohesion in dimers and trimers 
also affects their evaporation and condensation coefficients.

\bigskip\noindent{\bf IV. EVAPORATION OF LENNARD-JONES FLUIDS}
\bigskip

\noindent {\bf A. Monomers}
\bigskip

\begin{figure}[htb]
\centering
\includegraphics[width=2.5in]{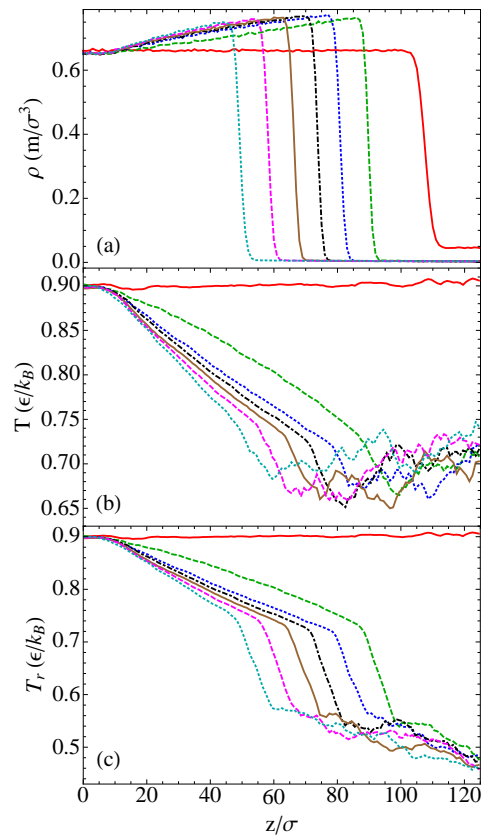}
\caption{(Color Online) (a) Density, 
(b) temperature [defined as the average kinetic energy, Eq.~(\ref{TempDef})], 
and (c) temperature [with mean velocity subtracted, Eq.~(\ref{TempDefRelative})] 
for a LJ monomer fluid at $T_b=0.9\epsilon/k_{\rm B}$ in contact with a vacuum. 
From right to left the profiles are plotted every $2000\tau$ 
since the evaporation process was started at $t=0$ (the rightmost curve).}
\label{EvapMonomerVac}
\end{figure}

When placed in contact with a vacuum, 
the evaporation of the LJ monomer system occurs very rapidly, 
as shown in Fig.~\ref{EvapMonomerVac}(a). 
The vapor is quickly depleted and the vapor density drops 
by an order of magnitude from its equilibrium value.
On the other hand, the liquid density increases 
by about $20\%$ from its equilibrium value near the interface \cite{{holyst09}} 
as there is significant evaporative cooling of the liquid film 
[Figs.~\ref{EvapMonomerVac}(b) and \ref{EvapMonomerVac}(c)], no matter the temperature
is measured either as $T(z)$ in Eq.~(\ref{TempDef}) 
or as $T_r(z)$ in Eq.~(\ref{TempDefRelative}). 
Note that there is essentially no advective motion in the liquid region,
and thus $\overline{v}=0$ and $T(z)=T_r(z)$ there.
However, since the vapor is very dilute when in contact with a vacuum,
all molecules leaving the liquid phase almost move freely toward 
the deletion zone and get removed. 
As a consequence of the absence of collision, 
the vapor region is far from local equilibrium and the mean velocity 
$\overline{v}$ is of order $\sqrt{k_{\rm B}T/m}$. 
Therefore $T(z)$ is very different from $T_r(z)$ in the vapor phase, 
as shown in Figs.~\ref{EvapMonomerVac}(b) and \ref{EvapMonomerVac}(c).

Though $T(z)$ and $T_r(z)$ are quite different in 
the vapor phase when in contact with a vacuum, 
they are the same in the liquid region in all cases.
They are also very close in both the liquid and vapor regions
in our other simulations described later 
where the evaporation rate is controlled and the vapor density during evaporation 
remains comparable to its equilibrium value. 
From an experimental perspective, 
it is the average kinetic energy that is measured
when a thermocouple is placed in the vapor phase during evaporation.
With these considerations, we refer to $T(z)$
as {\it temperature} in this paper and report its
measurements hereafter. 
Holyst and Litniewski\cite{holyst09}
called this temperature the  {\it pseudotemperature} and
concluded that it is very useful in
describing the evaporation kinetics.

Figure~\ref{EvapMonomerVac} shows that 
both the density and temperature gradients increase 
as the evaporation process continues. 
After a short time, the temperature has decreased across the interface 
from its initial value of $0.9\epsilon/k_{\rm B}$ to around $0.7\epsilon/k_{\rm B}$, 
which is very close to the triple point of this system.\cite{johnson93}
Similar results were found for $T_b=0.8\epsilon/k_{\rm B}$ and $1.0\epsilon/k_{\rm B}$, 
where the temperature at the liquid/vapor interface also drops to $\sim 0.7\epsilon/k_{\rm B}$. 
The large increase in density and decrease in temperature near the interface 
reported here for LJ monomers 
are much stronger than those of most common fluids such as water. 
In this sense, the particular feature of LJ model 
makes the effect of evaporation more dramatic.
The density enhancement near an evaporating interface was also
seen in the simulations of Holyst and Litniewski.\cite{holyst09}

\begin{figure}[htb]
\centering
\includegraphics[width=2.5in]{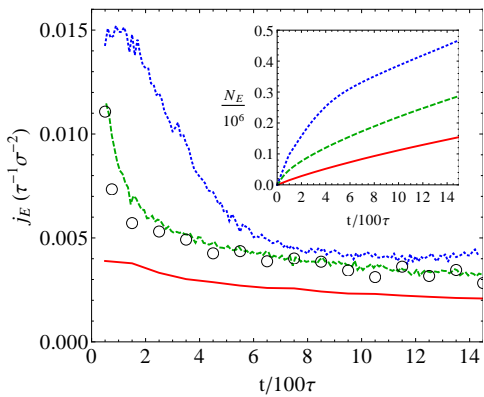}
\caption{(Color Online) Evaporation rate $j_E$ vs. time for a monomer
system in contact with a vacuum at $T_b=0.8$ (solid line), 
$0.9$ (dashed line), and $1.0\epsilon/k_{\rm B}$ (dotted line). 
The predicted evaporation flux $j_m/m$ 
based on Eq.~(\ref{MassFluxEq}) is shown as circles.
The inset shows the total number of removed atoms $N_E$ vs. time.}
\label{EvapRateMonomer}
\end{figure}

An evaporation rate $j_E$ is defined as the number of atoms removed 
in the deletion zone per unit time and area. 
It is related to the total number $N_E$ of removed atoms through 
\begin{equation}
j_E=\frac{1}{L_x L_y}\frac{{\rm d}N_E}{{\rm d}t}.
\label{EvapRateEq}
\end{equation}
Results for $j_E$ and $N_E$ are shown in Fig.~\ref{EvapRateMonomer} 
for LJ monomers at three temperatures.
As expected, $j_E$ is larger at higher $T_b$ 
since evaporation is a thermally activated process 
and occurs more rapidly at higher temperature.
The rate $j_E$ also shows strong time dependence.
In all cases, $j_E$ initially has a high value 
since the LJ fluid has a high vapor density 
and there are plenty of vapor molecules in the deletion zone. 
As the vapor is rapidly depleted, $j_E$ drops significantly over time 
and eventually reaches a plateau.
The final reduction factor, defined as the ratio between the value of $j_E$
at $t=0$ when the evaporation was initiated and that at very large $t$, 
is about $2$ for $T_b=0.8\epsilon/k_{\rm B}$ and 
increases to about $4$ for $T_b=0.9\epsilon/k_{\rm B}$ and $1.0\epsilon/k_{\rm B}$.

Holyst and Litniewski demonstrated that 
during evaporation of a liquid film into a vacuum the momentum flux,
$j_p\equiv n_{\rm vap} M \langle u_z^2 \rangle$ in the vapor phase far from the
liquid/vapor interface is equal to the pressure 
in the liquid film, $p_{\rm liq}$.\cite{holyst09} 
Here $n_{\rm vap}$ is the vapor number density, $M$ is the molecular mass, and
$\langle u_z^2 \rangle$ is the mean squared $z$ component of the molecular velocity.
From this observation they proposed an equation for the mass flux during
evaporation:
\begin{equation}
j_m \equiv n_{\rm vap} M \langle u_z \rangle = p_{\rm liq} 
\frac{\langle u_z \rangle}{\langle u_z^2 \rangle},
\label{MassFluxEq}
\end{equation}
where $\langle u_z \rangle$ is the mean $z$ component of the molecular velocity.
Note that $j_m$ has the unit of $mj_E$.
In our simulations, $p_{\rm liq}$, 
$\langle u_z \rangle$, and $\langle u_z^2 \rangle$ are averaged
in a thin region from $z_{\rm int}(t)+z_1$ to $z_{\rm int}(t)+z_2$, 
where $z_{\rm int}(t)$ denotes the location of interface at time $t$.
Typically, $z_1=-30\sigma$ and $z_2=-10\sigma$ for $p_{\rm liq}$, 
and $z_1=20\sigma$ and $z_2=40\sigma$ for 
$\langle u_z \rangle$ and $\langle u_z^2 \rangle$.
However, results are not sensitive 
to the exact location of these regions and varies by a few percent at most 
as long as they are far from the liquid/vapor interface.
The calculated mass flux $j_m$ from the above
equation at $T_b=0.9\epsilon/k_{\rm B}$ is shown in Fig.~\ref{EvapRateMonomer}. 
Clearly, the measured $j_E$ agrees with $j_m/m$ at most times, except
at the beginning stage of evaporation 
when $j_m/m$ tends to be a few percent smaller than $j_E$.

\begin{figure}[htb]
\centering
\includegraphics[width=3.25in]{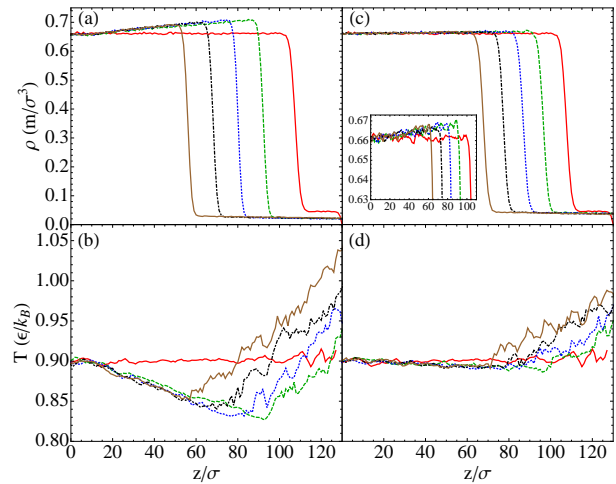}
\caption{(Color Online) Density [(a) and (c)] and temperature [(b) and (d)] for a LJ monomer 
fluid at $T_b=0.9\epsilon/k_{\rm B}$ evaporating at a rate 
$j_E=1.0\times 10^{-3}\tau^{-1}\sigma^{-2}$ [(a) and (b)] or 
$j_E=2.5\times 10^{-4}\tau^{-1}\sigma^{-2}$ [(c) and (d)]. 
From right to left the profiles are plotted 
every $8000\tau$ [(a) and (b)] or $24,000\tau$ [(c) and (d)]
since the evaporation process was started at $t=0$ (the rightmost curve).
The inset in (c) shows a blow up of the density in the interfacial region.}
\label{EvapMonomerEvapRate}
\end{figure}

We also ran simulations in which $j_E$ was controlled
by limiting the number of atoms removed from the deletion zone during a given simulation period.
Results for the density and temperature profiles are shown in 
Fig.~\ref{EvapMonomerEvapRate}
for two rates that are smaller than the final evaporation rate 
in the case of contacting with a vacuum, the plateau in Fig.~\ref{EvapRateMonomer},
by a factor around $4$ and $16$, respectively.
In Figs.~\ref{EvapMonomerEvapRate}(a) and \ref{EvapMonomerEvapRate}(b), 
the density enhancement and temperature drop
are still clearly visible, but the magnitudes are greatly reduced compared to 
those in Fig.~\ref{EvapMonomerVac}.
The vapor density in this case is also reduced from the equilibrium value but remains finite.
For a very slow evaporation rate 
[Figs.~\ref{EvapMonomerEvapRate}(c) and \ref{EvapMonomerEvapRate}(d)], 
these two effects almost disappear and the vapor density is close to the equilibrium density. 

In addition to the evaporative cooling, 
the temperature profiles shown here 
[Figs.~\ref{EvapMonomerVac}(b), \ref{EvapMonomerEvapRate}(b) 
and \ref{EvapMonomerEvapRate}(d)] 
indicate that the temperature of vapor 
increases with the distance from the liquid/vapor interface. 
This can be understood by a simple argument. 
molecules with higher kinetic energies evaporate faster and
more energetic molecules accumulate near the deletion zone, leading to a 
higher temperature at this end and a density gradient in the vapor phase.
Therefore, if the vapor temperature is measured at a certain distance 
away from the liquid/vapor interface, it will be higher than the actual temperature 
at the interface. 
This phenomenon is qualitatively consistent with the experimental measurement 
by Fang and Ward.\cite{fang99a}
They measured the temperature as close as one mean free path of the interface of
an evaporating liquid and the results indicated that 
the vapor temperature is indeed greater than that in the liquid phase at the interface. 
They also concluded this is due to more energetic molecules 
that are likely to evaporate first.
Note that the experimental condition in Ref.~\onlinecite{fang99a} 
is close to the situation where
the evaporation rate is controlled here. In this case, the vapor density 
is generally large enough to make local thermal equilibrium approximately true.
Our simulations confirmed that the mean velocity $\overline{v}$ 
is much less than $\sqrt{k_{\rm B}T/m}$, 
which leads to $T(z) \simeq T_r(z)$. 
Since the quantity recorded by the thermocouple 
is related to the average kinetic energy of vapor molecules,
the qualitative comparison made here between simulations and experiments 
should be reasonable.

\bigskip \noindent{\bf B. Dimers and Trimers}
\bigskip

While the LJ potential still serves as a reasonable approximation of 
inter-molecular interactions, 
most liquids of interest are composed of molecules not single atoms. 
One simple extension is to consider molecules of
two or more LJ monomers bound together. 
In the present simulations, the bonded interaction is realized 
through the introduction of a FENE potential [Eq.~(\ref{FENEPotential})] 
between bonded monomers.
The nonbonded interatomic interactions are still given 
by the LJ potential [Eq.~(\ref{LJPotential})].

\begin{figure}[htb]
\centering
\includegraphics[width=3.25in]{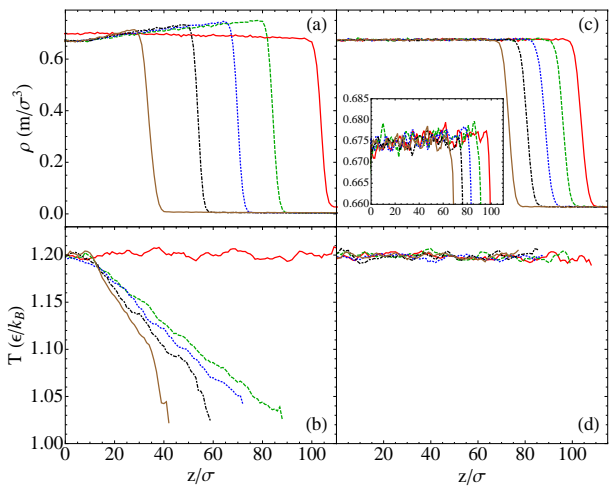}
\caption{(Color Online) Density [(a) and (c)] and temperature [(b) and (d)] for the LJ dimer 
fluid at $T_b=1.2\epsilon/k_{\rm B}$ in contact with a vacuum [(a) and (b)] or 
evaporating at a controlled rate 
$j_E=5.0\times 10^{-5}\tau^{-1}\sigma^{-2}$ [(c) and (d)]. 
From right to left the profiles are plotted every $5000\tau$ [(a) and (b)] 
and $100,000 \tau$ [(c) and (d)]
since the evaporation process was started at $t=0$ (the rightmost curve).
The inset in (c) blows up the density profile in the liquid and interfacial region.
Temperature in the vapor region is not included because the density of vapor is 
small and the data are too noisy to indicate a clear trend.}
\label{EvapDimerVacRate}
\end{figure}

\begin{figure}[htb]
\centering
\includegraphics[width=2.5in]{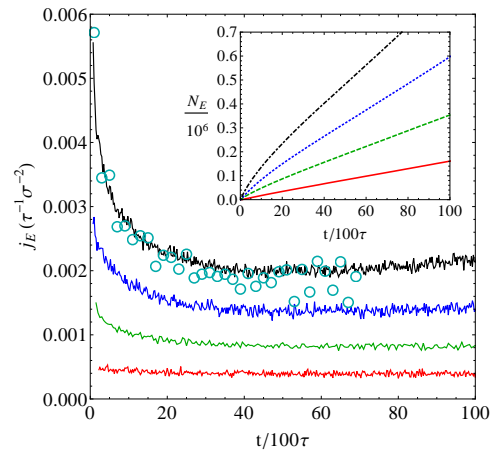}
\caption{(Color Online) Evaporation rate $j_E$ vs. time for the LJ dimer
system in contact with a vacuum at $T_b=0.9$, 
$1.0$, $1.1$, and $1.2\epsilon/k_{\rm B}$ (from bottom to top). 
The predicted evaporation flux $j_m/m$ 
based on Eq.~(\ref{MassFluxEq}) is shown as circles.
The inset shows the total number of removed atoms $N_E$ vs. time.}
\label{EvapRateDimer}
\end{figure}

\begin{figure}[htb]
\centering
\includegraphics[width=3.25in]{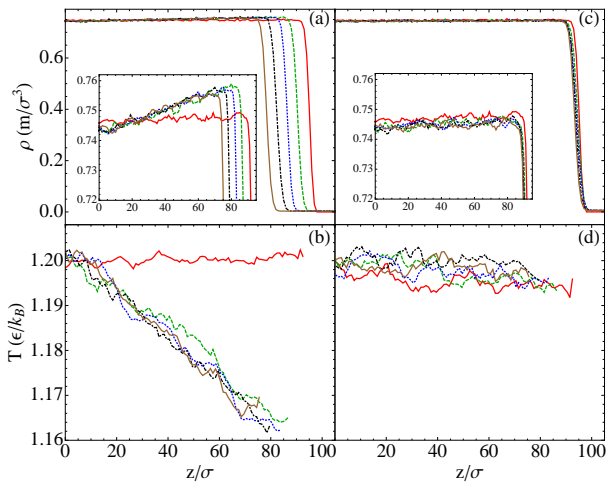}
\caption{(Color Online) Density [(a) and (c)] and temperature [(b) and (d)] for the LJ trimer
fluid at $T_b=1.2\epsilon/k_{\rm B}$ evaporating at a rate 
$j_E=5.0\times 10^{-4}\tau^{-1}\sigma^{-2}$ [(a) and (b)] or
$j_E=5.0\times 10^{-5}\tau^{-1}\sigma^{-2}$ [(c) and (d)]. 
From right to left the profiles are plotted every $60,000\tau$ in both cases
since the evaporation process was started at $t=0$ (the rightmost curve).
The insets in (a) and (c) show a blow up of density in the liquid and interfacial region.}
\label{EvapTrimerEvapRate}
\end{figure}

\begin{figure}[htb]
\centering
\includegraphics[width=2.5in]{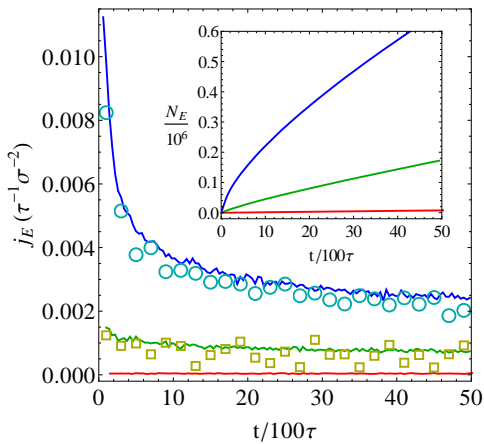}
\caption{(Color Online) Evaporation rate $j_E$ vs. time for the LJ trimer
system in contact with a vacuum at $T_b=0.9$, 
$1.2$, and $1.5\epsilon/k_{\rm B}$ (lines from bottom to top).
Open squares and circles show the predicted evaporation flux $j_m/m$
from Eq.~(\ref{MassFluxEq}) at
$T_b=1.2$ and $1.5\epsilon/k_{\rm B}$, respectively.
The inset shows the total number of removed atoms $N_E$ vs. time.}
\label{EvapRateTrimer}
\end{figure}

Results on the density and temperature profiles 
for the LJ dimer system in contact with a vacuum or evaporating at 
a controlled evaporation rate are shown in 
Fig.~\ref{EvapDimerVacRate} for $T_b=1.2\epsilon/k_{\rm B}$. 
Those for the trimer system at the same 
bulk temperature but at two controlled rates 
are shown in Fig.~\ref{EvapTrimerEvapRate}.
Results on the evaporation rate $j_E$ at various temperatures 
when in contact with a vacuum are shown
in Fig.~\ref{EvapRateDimer} (dimer) and Fig.~\ref{EvapRateTrimer} (trimer).

The phenomenology of evaporation into a vacuum for 
dimers [Fig.~\ref{EvapDimerVacRate}(a) and \ref{EvapDimerVacRate}(b)] 
and trimers (not shown) 
is similar to that of monomers, including the density enhancement and
the evaporative cooling near the liquid/vapor interface. 
However, the relative magnitudes of density increases and temperature drops 
are smaller for dimers and even smaller for trimers.
This is consistent with the observation 
that cohesion gets stronger and 
evaporation slows down in dimer and trimer fluids.

More quantitative information of evaporation into a vacuum 
is obtained by measuring the evaporation rate $j_E$.
It generally decreases with time and only approaches a constant value after a certain interval. 
Compared with monomers the reduction factor is smaller for dimers.
For the trimer system the time-dependence is weak 
except near the critical temperature $T_c$.
This is understandable because the vapor density decreases as the chain length increases. 
As expected, $j_E$ is higher at higher $T_b$.
At the same bulk temperature $T_b$, $j_E$ is largest for the monomer system and
smallest for the trimer.
For example, at $T_b=0.9\epsilon/k_{\rm B}$ 
it is around $2\times 10^{-3}\tau^{-1}\sigma^{-2}$ for monomers, 
$4\times 10^{-4}\tau^{-1}\sigma^{-2}$ for dimers, and 
$4\times 10^{-5}\tau^{-1}\sigma^{-2}$ for trimers. 
Thus increasing the number of atoms in a molecule only from $1$ to $3$ 
results in a decrease of more than a factor of $50$ in $j_E$.

The results for $j_m/m$ from Eq.~(\ref{MassFluxEq}) for dimers at 
$T_b=1.2\epsilon/k_{\rm B}$ and for trimers at 
$T_b=1.2\epsilon/k_{\rm B}$ and $1.5\epsilon/k_{\rm B}$ are shown
in Figs.~\ref{EvapRateDimer} and \ref{EvapRateTrimer}, respectively.
In all cases, the agreement between $j_m/m$ and $j_E$ is quite satisfactory, 
indicating that Eq.~(\ref{MassFluxEq}) is valid not only for monatomic liquids, 
but also for molecular liquids such as dimers and trimers.
We also note that in evaluating $\langle u_z^2 \rangle$ in Eq.~(\ref{MassFluxEq}),
molecular velocities should be used. 

The magnitude of the density enhancement and temperature drop near the interface
also depends on the evaporation rate. 
As shown in Fig.~\ref{EvapDimerVacRate}(c) and (d), 
for the dimer system at $T_b=1.2\epsilon/k_{\rm B}$ they are not observed 
when $j_E$ is reduced to $5.0\times 10^{-5}\tau^{-1}\sigma^{-2}$,
which is much smaller than the plateau value of $j_E$ of this 
system in contact with a vacuum.
For the trimer system under the same temperature, 
the density and temperature gradients are visible at 
$j_E=5.0\times 10^{-4}\tau^{-1}\sigma^{-2}$ 
[Fig.~\ref{EvapTrimerEvapRate}(a) and \ref{EvapTrimerEvapRate}(b)], 
but disappear when
$j_E$ is reduced further by a factor of $10$ to 
$j_E=5.0\times 10^{-5}\tau^{-1}\sigma^{-2}$ 
[Fig.~\ref{EvapTrimerEvapRate}(c) and \ref{EvapTrimerEvapRate}(d)].
Note that the plateau value of $j_E$ for trimers
at this temperature when evaporating into a vacuum is 
$7.7\times 10^{-4}\tau^{-1}\sigma^{-2}$.
Thus it is reasonable to see density and temperature gradients
at $j_E=5.0\times 10^{-4}\tau^{-1}\sigma^{-2}$, 
but not at much slower evaporation rates.

\bigskip \noindent{\bf C. Stagnation Pressure During Evaporation}
\bigskip

During evaporation, the vapor escaping into a vacuum is far from equilibrium and 
a local temperature and pressure defined with respect to
a thermodynamic equilibrium cannot be used for its description.
Holyst and Litniewski showed for LJ monomers that global quantities,
such as total kinetic energy, total mass flux, and total momentum flux,
can be well defined and provide useful information about the
physical state of the vapor phase.\cite{holyst09} Particularly, 
a stagnation pressure tensor can be defined as
\begin{equation}
p_{\alpha\beta}=\frac{1}{V}\left( \sum_j m_j v_{j\alpha} v_{j\beta} -
\sum_{i>j} \sum_j r_{ij\alpha} \frac{\partial \phi_{ij}}{\partial r_{ij\beta}} 
\right),
\label{stagEq}
\end{equation}
where $i$ and $j$ index all atoms inside the volume $V$, 
$m_j$ and $v_j$ are the atomic mass and velocity,
$r_{ij}$ and $\phi_{ij}$ are the interatomic distance and potential, respectively, 
and Greek indices indicate components along the $x$, $y$, and $z$ directions.
Note that this pressure tensor is not defined relative to a local equilibrium.
Holyst and Litniewski demonstrated that for the stagnation pressure 
the $p_{zz}$ component in the vapor phase 
is equal to the liquid pressure, $p_{\rm liq}$, far from the interface.\cite{holyst09} 

\begin{figure}[htb]
\centering
\includegraphics[width=2.5in]{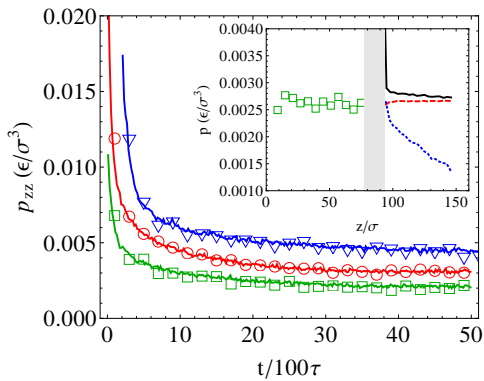}
\caption{(Color Online) Stagnation pressure $p_{zz}$ averaged in the 
liquid phase (symbols) and vapor phase (lines) for: 
monomers at $T_b=0.9\epsilon/k_{\rm B}$ (the middle line and $\bigcirc$);
dimers at $T_b=1.2\epsilon/k_{\rm B}$ (the bottom line and $\Box$); 
and trimers at $T_b=1.5\epsilon/k_{\rm B}$ (the top line and $\bigtriangledown$).
The trimer data are shifted upward by $0.002\epsilon/\sigma^3$ 
and right by $200\tau$ for clarity.
The inset shows the stagnation pressure $p$ as a function 
of distance along the normal direction of the 
liquid/vapor interface for trimers at $T_b=1.5\epsilon/k_{\rm B}$. 
The pressure in the 
liquid phase, $p_{\rm liq}$, is shown as squares ($\Box$), and 
that in the vapor phase, $p_{\rm vap}$, 
is shown as the dashed line (the $p_{zz}$ component)
and the dotted line [the $(p_{xx}+p_{yy})/2$ component].
The solid line represents the momentum flux 
$j_p\equiv n_{\rm vap} M \langle u_z^2 \rangle$.
The interfacial region is denoted as a shadowed strip.
Data are for systems evaporating into a vacuum.}
\label{StagPressFig}
\end{figure}

We measured the stagnation pressure for Lennard-Jones monomers, dimers,
and trimers. Results are shown in Fig.~\ref{StagPressFig}.
Indeed, for all systems at all times the $p_{zz}$ component in the vapor phase 
(solid lines in the main panel of Fig.~\ref{StagPressFig})
is equal to the pressure in the liquid film (symbols). 
Data in Fig.~\ref{StagPressFig} are for liquid films evaporating into a vacuum.
We confirmed that the equality is also true for systems evaporating at
controlled rates.

The inset of Fig.~\ref{StagPressFig} shows the stagnation pressure
along the normal direction of the liquid/vapor interface. 
In the liquid phase, $p_{zz}=p_{xx}=p_{yy}=p_{\rm liq}$.
However, in the vapor phase, $p_{zz}\ne p_{xx}=p_{yy}$, 
and the difference grows as the distance away from the interface.
The inset also shows that the momentum flux, 
$j_p=n_{\rm vap}M\langle u_z^2 \rangle$, approaches
$p_{zz}$ in the vapor phase far from the interface. 
This is not surprising for monatomic systems since in this case 
$j_p$ is exactly the first term in the expression for $p_{zz}$ 
in Eq.~(\ref{stagEq}). The second term in $p_{zz}$ represents
the contribution from interatomic interactions. 
For systems evaporating into a vacuum, the vapor becomes
increasingly dilute away from the liquid/vapor interface
so that the contribution to $p_{zz}$ from interactions between two monomers 
essentially vanishes and the momentum flux dominates.
For molecular systems, $j_p$ is not identical
to the first term in $p_{zz}$ as expressed in Eq.~(\ref{stagEq}), 
which is based on atomic velocities not molecular velocities averaged
from those of its constitute atoms. However,
$p_{zz}$ can be also defined 
in terms of molecular velocities and the 
interactions between molecules instead of atoms.\cite{allen87}
Then molecular density becomes dilute during
evaporation into a vacuum and the same argument
above applies again. The result is that
$j_p$ still dominates over the interaction contribution in $p_{zz}$ 
in molecules escaping into a vacuum.
This relation between $p_{zz}$ and $j_p$ in a dilute vapor
is the foundation of Eq. (\ref{MassFluxEq}), 
in which the mass flux $j_m$ can be determined from $p_{\rm liq}$
since $p_{\rm liq}$ equals $p_{zz}$ in the vapor phase. 
As shown in Figs.~\ref{EvapRateMonomer}, \ref{EvapRateDimer}, 
and \ref{EvapRateTrimer}, the agreement between $j_m/m$ and 
the evaporation rate $j_E$ is very good 
for all systems evaporating into a vacuum.

For systems evaporating at small controlled rates, the vapor density
is comparable to its equilibrium value and it is not expected 
that $j_p$ is equal to $p_{zz}$ even in the vapor phase, 
though $p_{zz}$ is still equal to $p_{\rm liq}$. 
As a result, the mass flux $j_m$ from Eq.~(\ref{MassFluxEq})
is generally far from the evaporation rate $mj_E$. 
A simulation of LJ monomers evaporating at 
$j_E=1.0\times 10^{-3}\tau^{-1}\sigma^{-2}$ and $T_b=0.9\epsilon/k_{\rm B}$ 
shows that $j_m/m$ from Eq.~(\ref{MassFluxEq}) is only about $50\%$ of
the measured $j_E$.

\bigskip\noindent{\bf V. COMPARISON TO KINETIC THEORY}
\bigskip

At the liquid/vapor interface, the molecular exchange between the two phases 
occurs continuously in the form of evaporation and condensation.
Results discussed above show that the molecular composition 
affects the evaporation rate to a large extent.
It remains an interesting question if the change
in molecular composition also affects 
the atomic kinetics during evaporation and condensation, 
such as velocity distributions of the molecules.
In equilibrium, all molecules leaving the interface to the gas phase 
or arriving at the interface from the gas phase 
have a normal velocity $v_z$ that
satisfies the MB distribution 
with a probability density function
\begin{equation}
p(v_z)=\frac{Mv_z}{k_{\rm B}T_i} {\rm exp}\left( -\frac{M v_z^2}{2 k_{\rm B}T_i}\right),
\label{MBDist}
\end{equation}
where $M$ is the molecular mass and $T_i$ is the temperature at the interface.
Note that $T_i$ equals $T_b$ in equilibrium 
but is generally lower than $T_b$ when there is net evaporation.

\begin{figure}[htb]
\centering
\includegraphics[width=3.25in]{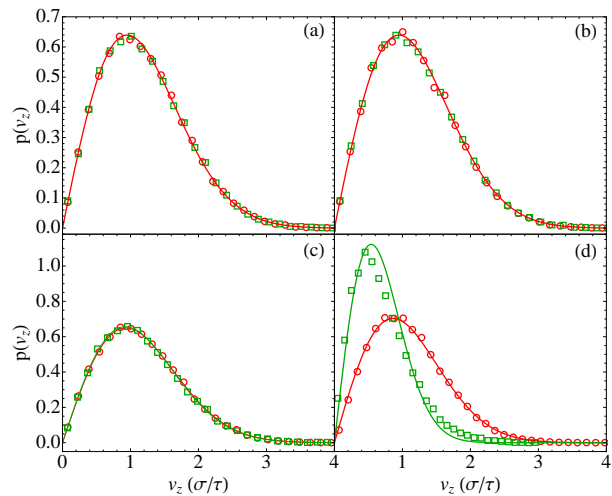}
\caption{(Color Online) Velocity distributions of $v_z$ for 
all molecules that leave (circles) and arrive (squares) at the interface for
monomer systems at $T_b=0.9\epsilon/k_{\rm B}$ under various evaporation conditions: 
(a) liquid/vapor equilibrium; 
(b) controlled rate $j_E=2.5\times 10^{-4}\tau^{-1}\sigma^{-2}$;
(c) controlled rate $j_E=1.0\times 10^{-3}\tau^{-1}\sigma^{-2}$; 
(d) in contact with a vacuum.
Solid lines in (a) and (b) represent 
the MB distribution with $T_i=T_b=0.9\epsilon/k_{\rm B}$.
Two solid lines in (c) represent 
the MB distribution with $T_i=0.871\epsilon/k_{\rm B}$ for molecules that leave the interface 
and $T_i=0.854\epsilon/k_{\rm B}$ for molecules that arrive at the interface. 
Two solid lines in (d) represent 
the MB distribution with $T_i=0.744\epsilon/k_{\rm B}$ for molecules that leave the interface 
and $T_i=0.292\epsilon/k_{\rm B}$ for molecules that arrive at the interface.}
\label{EvapCondVelDist}
\end{figure}

To analyze the velocity distributions, we consider a plane 
just outside the liquid/vapor transition layer. 
All molecules crossing this plane and entering (leaving) the 
transition layer are counted as incoming (outgoing) molecules. 
Their normal velocity $v_z$ is measured and the 
corresponding distribution function $p(v_z)$ is calculated.
The results for LJ monomers are shown in Fig.~\ref{EvapCondVelDist}. 
Those for dimers and trimers are very similar and are not shown.
The results are not sensitive to the position of 
the plane that molecules cross as long as it is outside 
but not far, within several $\sigma$, from the transition layer.

As expected, Fig.~\ref{EvapCondVelDist}(a) shows that 
when the liquid/vapor interface is in equilibrium 
the MB distribution is satisfied 
by molecules that enter or exit the transition layer
with a $T_i$ that is the same as the bulk temperature 
in the liquid and vapor phases. 
This is also the case when the evaporation rate is as low as 
$2.5\times 10^{-4}\tau^{-1}\sigma^{-2}$ [Fig.~\ref{EvapCondVelDist}(b)].
When the evaporation rate is increased to $1.0\times 10^{-3}\tau^{-1}\sigma^{-2}$ 
[Fig.~\ref{EvapCondVelDist}(c)],
the MB distribution still holds, but corresponds to a $T_i$ 
that is slightly lower than $T_b$, which is fixed by a thermostat 
near the lower confining wall.
Furthermore, in this case slightly different $T_i$'s have to be 
used for molecules that enter or exit the transition layer.

The situation is quite different in the case of evaporation into a vacuum,
where the evaporation rate is very high. 
Figure \ref{EvapCondVelDist}(d) shows that the velocity distributions 
of molecules entering or exiting the transition layer have 
very different $T_i$'s. 
For the outgoing molecules, $T_i=0.744\epsilon/k_{\rm B}$ 
is consistent with the temperature of the 
liquid/vapor interface, which is lower than the bulk temperature 
$T_b=0.9\epsilon/k_{\rm B}$ of liquid far from the interface.
This reduction of the temperature at the liquid/vapor interface 
is due to evaporative cooling. 
For the arriving molecules we found $T_i=0.292\epsilon/k_{\rm B}$ 
when $p(v_z)$ was fit to a MB distribution.
This temperature is much lower than that for the outgoing molecules 
and is only about $1/3$ of $T_b$.
The strong asymmetry between $T_i$'s for the outgoing and incoming molecules
can be understood as follows.
When the liquid evaporates into a vacuum, 
the vapor density is extremely low and it is very unlikely
that a vapor molecule undergoes a collision with other molecules in the vapor phase and 
is reflected back to the liquid/vapor interface. 
All molecules leaving the interface tend to move 
ballistically toward the deletion zone and get removed. 
This is particularly the case for outgoing molecules 
with high velocities $v_z$ normal to the interface because 
they would require multiple collisions to be reflected, which is very rare.
Therefore, most reflected molecules are those moving slowly.
This leads to a strong imbalance in the distributions 
of $v_z$ of the incoming and outgoing molecules, 
which is manifest in the fact that the incoming molecules have 
a much lower $T_i$ than that of the outgoing molecules.

Results in Fig.~\ref{EvapCondVelDist}(d) indicate that 
if the full velocity distribution
is plotted for all the molecules approaching or 
leaving the liquid/vapor interface, one would expect some
deviation from the MB distribution that is symmetric between $v_z$ and $-v_z$. That is,
the contribution from positive values of $v_z$ have more weight 
than that on the negative side.
This deviation was also observed in previous MD simulations 
of evaporation into a vacuum.\cite{anisimov97,frezzotti05}
Our results further indicate that this asymmetry also occurs 
when the evaporation rate is high enough that 
the vapor density is substantially reduced from its equilibrium value.

Not all molecules arriving at the liquid/vapor interface 
from the gas phase condense into the liquid phase. 
A fraction is reflected back into the vapor phase.
The condensation coefficient defined in Introduction quantifies this effect.
It essentially represents the fraction of
incoming molecules that indeed become attached to the liquid phase.
It has been argued that the condensation coefficient is close to unity 
for monatomic liquids such as liquid metals,\cite{niknejad81}
and less than unity for polyatomic liquids because rotational motion of 
polyatomic molecules in the liquid state make it more difficult 
to accommodate newly arriving molecules.\cite{fujikawa90}

The reflected molecules obviously contribute to the evaporation flux.
Equivalently, not all evaporated molecules come directly from the liquid phase.
The evaporation coefficient is 
the ratio between the molecular flux due to real evaporation, 
i.e., from those molecules transformed into vapor from the liquid phase, 
and the maximum flux $j_{\rm max}$ calculated from the MB distribution. 
It is easy to derive the HK formula, $j_{\rm max}=\frac{1}{4}n\overline{v}$, 
where $n$ is the density of the saturated vapor corresponding to $T_i$
and $\overline{v}$ is the mean molecular velocity.
For liquid/vapor equilibrium the condensation 
and evaporation coefficients must be the same since 
the velocity of molecules arriving at the interface also satisfies
the MB distribution.

Hereafter, we consider the total evaporation flux as the sum of 
that due to the true evaporation of liquid molecules and 
another due to reflection of incoming vapor molecules.
Results in Fig.~\ref{EvapCondVelDist}(d) show that 
when a liquid evaporates into a vacuum, 
the reflection flux is greatly suppressed because of the depletion of
vapor. In this case the total evaporation flux is very close to
the true evaporation flux. 
This is the reason that Ishiyama {\it et al.} emphasized that 
the condensation coefficient can be determined without any ambiguity
by measuring the spontaneous evaporation flux from the simulations
of evaporation into a vacuum.\cite{fujikawa04}

\begin{figure}[htb]
\centering
\includegraphics[width=2.5in]{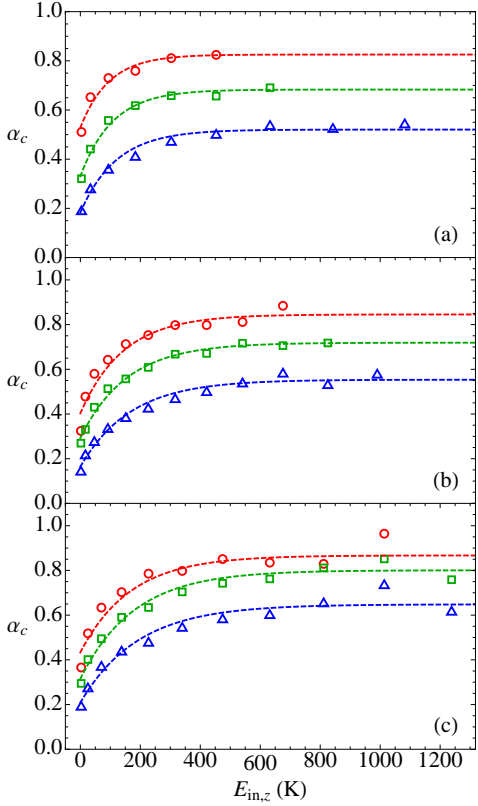}
\caption{(Color Online) Condensation probability $\alpha_c$ as a function
of $E_{{\rm in},z}$, which is 
the normal component of the translational energy of incoming molecules.
Data presented here are measured on liquid/vapor equilibrium systems 
at various temperatures.
However, results are similar if the evaporation rate is fixed at a small value and 
the liquid/vapor interface is near equilibrium.
(a) monomers: $T_b=0.8\epsilon/k_{\rm B}$ (triangles), $0.9\epsilon/k_{\rm B}$ (squares), 
and $1.0\epsilon/k_{\rm B}$ (circles);
(b) dimers: $T_b=1.2\epsilon/k_{\rm B}$ (triangles), $1.3\epsilon/k_{\rm B}$ (squares), 
and $1.4\epsilon/k_{\rm B}$ (circles);
(c) trimers: $T_b=1.4\epsilon/k_{\rm B}$ (triangles), $1.5\epsilon/k_{\rm B}$ (squares), 
and $1.6\epsilon/k_{\rm B}$ (circles).
In this plot a conversion factor $\epsilon/k_{\rm B}=119.8{\rm K}$ (for argon) is used.
}
\label{CondProbPlot}
\end{figure}

Through MD simulations of LJ monomers, Tsuruta {\it et al.} discovered 
that the condensation coefficient is actually 
an average quantity.\cite{tsuruta99,tsuruta03,tsuruta05}
For incoming molecules with different normal translational energy 
$E_{{\rm in},z}=Mv_z^2/2$, their tendency to condensate into the liquid phase
is generally different. 
A condensation probability $\alpha_c$ 
was defined to quantify this tendency and they suggested that 
$\alpha_c$ depends on the normal velocity of incident molecules
in the following functional form,
\begin{equation}
\alpha_c = \alpha \left[ 1-\beta {\rm exp} \left( 
\frac{-E_{{\rm in},z}}{k_{\rm B}T_i} \right) \right],
\label{CondProbEq}
\end{equation}
where $\alpha$ and $\beta$ are two parameters that depend on the 
properties of liquid and temperature. 
In this framework, the condensation coefficient 
discussed above is the average of $\alpha_c$
over the velocity distribution of incoming molecules and 
is denoted as $\overline{\alpha_c}$.
Since the velocity of all incoming molecules satisfies a MB distribution, 
$\overline{\alpha_c}$ can be expressed as
\begin{equation}
\begin{array}{lll}
\overline{\alpha_c} & = & \int_{0}^{+\infty} \alpha_c (v_z) p(v_z) dv_z \\
& = & \alpha (1-\beta/2).
\end{array}
\label{AveCondCoeffEq}
\label{CondCoeffEq}
\end{equation}

In our simulations we grouped incoming molecules by the normal component $v_z$ of their velocity
and measured their condensation probability $\alpha_c$. 
However, an ambiguity arises about what incoming molecules should be counted as 
condensed ones since all molecules arriving at the transition layer 
will eventually leave after some period of time. 
We counted those incoming molecules that stay in the liquid phase 
or the liquid/vapor transition zone for at least $\Delta t=25\tau$. 
Since a typical value $\tau$ is approximately $2$ps,\cite{rapaport95} 
$\Delta t$ is about $50$ps.
This choice of $\Delta t$ is somewhat {\it ad hoc}. 
However, previous MD simulations have established that the characteristic time of energy
excitation during evaporation or relaxation during condensation
is roughly $50 - 70$ps.\cite{yasuoka94} 
For this reason our choice of condensation time scale as $50$ps should be reasonable.
This choice is further substantiated by the fact that 
the measured $\alpha_c$ does not change significantly even 
if $\Delta t$ is increased by a factor of $2$ to $50\tau$.
All data presented here are for $\Delta t=25\tau$.
After $\alpha_c$ was obtained, the $\alpha_c \sim v_z$ was fit to Eq.~(\ref{CondProbEq}) with 
$\alpha$ and $\beta$ as fitting parameters, 
from which the condensation coefficient 
$\overline{\alpha_c}$ was calculated using Eq.~(\ref{AveCondCoeffEq}).

Results for $\alpha_c (v_z)$ are shown in Fig.~\ref{CondProbPlot}.
It indicates that $\alpha_c$ indeed follows the functional form in Eq.~(\ref{CondProbEq}), 
regardless of the change in molecular composition from monomers to dimers and to trimers.
Vapor molecules arriving at the liquid/vapor interface with a small $v_z$ 
are less likely to condense and join the liquid state. 
This can be understood by a simple physical picture.
When the arriving molecules collide with liquid molecules, they are more likely to 
be reflected after one or several collisions if their velocity is small.
Arriving molecules with a large velocity $v_z$
are able to penetrate the liquid region deeper, and 
have a bigger chance to survive multiple collisions 
and to remain in the liquid phase for at least a period $\sim \Delta t$.

The condensation probability $\alpha_c$ being a function of $v_z$ has another implication.
Since all molecules leaving the liquid/vapor interface are either those 
that truly evaporate from the liquid
or those that have been reflected, 
and the velocity of all leaving molecules satisfies the MB distribution 
(as shown in Fig.~\ref{EvapCondVelDist}), 
the normal velocity distributions of the truly evaporated molecules and
reflected ones are modified from the MB distribution. 
Particularly, the MB distribution is modified by the condensation probability 
$\alpha_c$ for the truly evaporated molecules
and by $1-\alpha_c$ for the reflected ones. 
Thus they have the following forms after a proper normalization,\cite{tsuruta99}
\begin{equation}
p_e(v_z)=\frac{1-\beta {\rm exp}\left( -\frac{M v_z^2}{2 k_{\rm B}T_i}\right)}{1-\beta/2}
\frac{Mv_z}{k_{\rm B}T_i} 
{\rm exp}\left( -\frac{M v_z^2}{2 k_{\rm B}T_i}\right),
\label{ModMBDistEvap}
\end{equation}
and
\begin{equation}
p_r(v_z)=\frac{1-\alpha+\alpha\beta {\rm exp}\left( -\frac{M v_z^2}{2 k_{\rm B}T_i}\right)}
{1-\alpha+\alpha\beta/2}
\frac{Mv_z}{k_{\rm B}T_i} 
{\rm exp}\left( -\frac{M v_z^2}{2 k_{\rm B}T_i}\right),
\label{ModMBDistRefl}
\end{equation}
where the subscripts $e$ and $r$ stand for ``truly evaporated'' and ``reflected'', respectively.

\begin{figure}[htb]
\centering
\includegraphics[width=2.5in]{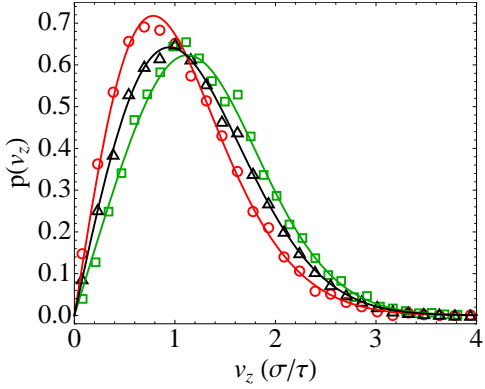}
\caption{(Color Online) Velocity distributions of $v_z$ of all molecules which 
left (triangles), were
evaporated (squares), or were reflected (circles) from the interface
for LJ monomers at $T_b=0.9\epsilon/k_{\rm B}$ evaporating at a rate  
$j_E=2.5\times 10^{-4}\tau^{-1}\sigma^{-2}$. 
Solid lines are fits to the MB distribution $p(v_z)$ in Eq.~(\ref{MBDist}) and
the modified MB distributions $p_e(v_z)$ and $p_r(v_z)$
in Eqs.~(\ref{ModMBDistEvap}) and (\ref{ModMBDistRefl}), respectively. }
\label{EvapReflectAll}
\end{figure}

The distribution functions $p_e(v_z)$ and $p_r(v_z)$ for the truly evaporated and reflected molecules
were measured directly in simulations and provided another way to determine $\alpha$ 
and $\beta$, and eventually the condensation coefficient $\overline{\alpha_c}$. 
One example is shown in Fig.~\ref{EvapReflectAll}, which includes data on 
$p(v_z)$, $p_e(v_z)$, and $p_r(v_z)$ of monomers evaporating at a small fixed rate.
The successful fits to the MB and modified MB distributions 
in Eqs.~(\ref{MBDist}), (\ref{ModMBDistEvap}), and (\ref{ModMBDistRefl}) 
indicate that the condensation probability $\alpha_c$ provides a reasonable quantitative measure
of the condensation and evaporation processes. 
Furthermore, we compared the condensation coefficient $\overline{\alpha_c}$
determined either using $\alpha_c(v_z)$ or using $p_e(v_z)$ and $p_r(v_z)$. 
For systems in equilibrium or not far from equilibrium 
such as evaporation with small controlled rates,
the two results generally agree, which further validates the 
kinetic model of evaporation described above.
For systems far from equilibrium such as evaporation into a vacuum, 
it is generally very difficult to measure $\overline{\alpha_c}$ in these ways
because there are essentially no incoming molecules and thus no reflected flux. 
In this case, it might be possible to determine $\overline{\alpha_c}$ 
by directly measuring the evaporation flux and comparing 
to the theoretical maximum flux $j_{\rm max}$, 
as suggested in Ref.~\onlinecite{fujikawa04}. 
However, the temperature of the liquid/vapor interface needs to be determined first 
in order to calculate $j_{\rm max}$. 
This is not an easy task as the interface is moving during evaporation and
the temperature is sensitive to the location of measurement in 
the region around the interface.

\begin{figure}[htb]
\centering
\includegraphics[width=3in]{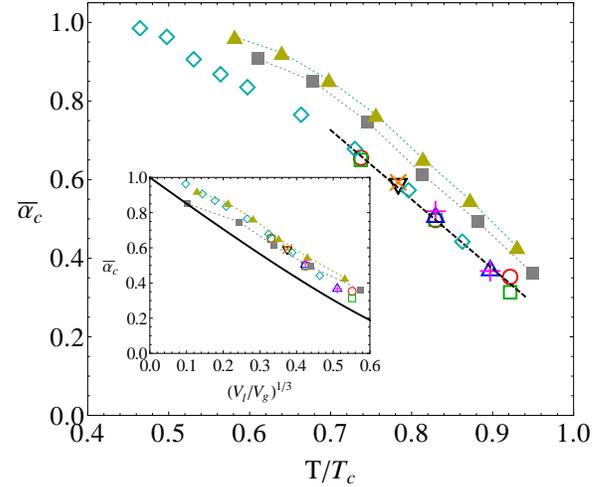}
\caption{(Color Online) Condensation coefficient $\overline{\alpha_c}$ vs. 
temperature for monomers, dimers,
and trimers under various evaporation conditions.
For monomers: circles ($\bigcirc$) and triangles ($\bigtriangleup$ and $\bigtriangledown$)
are measured using $\alpha_c(v_z)$; 
open squares ($\Box$), pluses ($+$), and crosses $\times$ 
are measured using $p_e(v_z)$ and $p_r(v_z)$;
circles ($\bigcirc$) and open squares ($\Box$) are 
for liquid/vapor equilibrium systems at $T_i=T_b=0.8\epsilon/k_{\rm B}$, 
$0.9\epsilon/k_{\rm B}$, and $1.0\epsilon/k_{\rm B}$; 
upward triangles ($\bigtriangleup$) and pluses ($+$) are for 
evaporation rate $j_E=2.5\times 10^{-4}\tau^{-1}\sigma^{-2}$ 
at $T_b=0.9\epsilon/k_{\rm B}$ and $1.0\epsilon/k_{\rm B}$; 
downward triangles ($\bigtriangledown$) and crosses ($\times$) are for 
evaporation rate $j_E=1.0\times 10^{-3}\tau^{-1}\sigma^{-2}$ 
at $T_b=0.9\epsilon/k_{\rm B}$.
For data measured during evaporation, 
the actual temperature $T_i$ of the liquid/vapor interface is used.
The data for dimers and trimers are shown with solid squares ($\blacksquare$) 
and solid triangles ($\blacktriangle$), respectively, 
and both are calculated 
using $\alpha_c(v_z)$ from liquid/vapor equilibrium systems.
The dashed line is a linear fit to the monomer data at high temperature.
The dotted lines are guides to the eye.
Our data of LJ monomers agree with those from 
Ref.~\onlinecite{fujikawa04} [shown with open diamonds ($\Diamond$)], 
which were determined by directly measuring the evaporation flux into a vacuum.
The inset plots $\overline{\alpha_c}$ as a function of the translational length ratio, 
$(V_l/V_g)^{1/3}$, between the vapor and liquid phase. 
The solid line in the inset is the prediction  
of the transition state theory (see Ref.~\onlinecite{tsuruta03}).}
\label{CondCoeffAll}
\end{figure}

Figure \ref{CondCoeffAll} shows results on 
the condensation coefficient $\overline{\alpha_c}$ 
for LJ monomers, dimers, and trimers. 
The agreement between results from the condensation probability
and the velocity distributions is clear. 
Generally, $\overline{\alpha_c}$ decreases with a increasing $T$.
For monomers in temperature range we could study, 
the reduction is approximately linear. 
The expected crossover in $\overline{\alpha_c}$ to unity 
for lower temperatures is not observed for monomers 
since for lower $T$ the system crystallizes.
But the crossover can be identified in the data from Ref.~\onlinecite{fujikawa04}, 
which used the Dymond-Alder potential for argon in the low temperature range.
For dimers and trimers, the expected crossover 
of $\overline{\alpha_c} \sim T$ at lower temperatures is evident.

Figure \ref{CondCoeffAll} shows that the value $\overline{\alpha_c}$ depends
on the evaporation condition. This is obvious if we examine the monomer data at 
$T_b=0.9\epsilon/k_{\rm B}$. 
The value of $\overline{\alpha_c}$ in liquid/vapor equilibrium is slightly
lower than its value at the evaporation rate $j_E=2.5\times 10^{-4}\tau^{-1}\sigma^{-2}$,
and both are clearly lower that the value at the larger evaporation rate 
$j_E=1.0\times 10^{-3}\tau^{-1}\sigma^{-2}$.
However, if we take into account the fact the temperature $T_i$ of the liquid/vapor interface
get lower when evaporation is stronger and use $T_i$ instead of $T_b$ for the horizontal axis, 
all data fall nicely onto the same master curve indicated 
by the straight line in Fig.~\ref{CondCoeffAll}. 
The reason that $\overline{\alpha_c}$ is larger under stronger evaporation 
is that the interface temperature decreases due to evaporative cooling and
lower temperatures corresponds to larger $\overline{\alpha_c}$.
This trend is also indicated by the data at $T_b=1.0\epsilon/k_{\rm B}$ 
for an equilibrium interface (thus $T_i=T_b$) and 
for a fixed evaporation rate $j_E=2.5\times 10^{-4}\tau^{-1}\sigma^{-2}$ (thus $T_i < T_b$).
In the latter $\overline{\alpha_c}$ is found larger.

Our results show that the condensation coefficient $\overline{\alpha_c}$ 
is substantially less than unity for LJ monomers above the triple-point temperature, 
and is higher for LJ dimers and trimers at the same reduced temperature $T/T_c$. 
This is in contrast to the traditional viewpoint that 
the condensation coefficients decreases when the molecular composition changes from
monatomic to polyatomic. In this viewpoint, the rotational motion of polyatomic molecules
acts as a constraint to make the vapor condensation less likely. 
However, the short chain molecules we have simulated are relatively flexible. 
When two polyatomic molecules come in contact with the interface, 
their attractive interaction is stronger
than that between two monatomic molecules 
because of the increase in the number of interacting sites. 
Furthermore, because of the flexibility, 
rotational degrees of freedom play a less important role.
This leads to stronger cohesion for polyatomic chain molecules. 
It is therefore not surprising that $\overline{\alpha_c}$ is 
higher for LJ dimers and trimers at the same $T/T_c$ because it basically
represents how likely a molecule can bind to its liquid phase.
Tsuruta and Nagayama also found that translational degrees of freedom 
dominate in the evaporation and condensation of water.\cite{tsuruta04}

Using transition state theory, 
Nagayama and Tsuruta derived a theoretical expression for $\overline{\alpha_c}$
in the case that the evaporation and condensation processes 
are dominated by translational motion of molecules.\cite{tsuruta03}
In this framework,
$\overline{\alpha_c}$ has a universal functional dependence on 
a translational length ratio, which is defined by the cubic root of 
the free volume ratio between the liquid and vapor phase, 
$(V_l/V_g)^{1/3}$.
The results for $\overline{\alpha_c}$ versus $(V_l/V_g)^{1/3}$ 
are shown in the inset of Fig.~\ref{CondCoeffAll}.
All data collapse onto a master curve except 
for dimers at low temperature. 
However all data are above the theoretical line.
Most data for $\overline{\alpha_c}$ in Ref.~\onlinecite{tsuruta03}
are also above the predicted 
curve for $\overline{\alpha_c}$ vs. $(V_l/V_g)^{1/3}$.

\bigskip\noindent{\bf VI. SUMMARY AND CONCLUSIONS}
\bigskip

We have used MD simulations to investigate fluids made of 
LJ monomers, dimers, and trimers.
The phase diagrams for the dimer and trimer systems 
were determined and their evaporation processes were simulated.
Our results show that a simple change from
monomer to dimer or trimer molecules has a strong effect on the evaporation rates.
The evaporation rate of monomers when in contact with a vacuum is extremely high 
and there are strong evaporative cooling 
and liquid density enhancement near the liquid/vapor interface.
All these effects are greatly reduced in dimer and trimer fluids. 
A physical explanation is provided on the basis that cohesion gets stronger for 
fluids made of longer chain molecules because of the increase in the number of 
interacting sites between two close molecules. 
The flexibility of the linear chain also suppresses the importance of 
rotational motion of molecules, which tends to reduce the cohesion.
The net outcome of the competition between these factors is that the surface tension
is higher and the evaporation slows down
at the liquid/vapor interface for fluids made of longer chains.

The measured evaporation rates for liquids evaporating into a vacuum
were compared to the modified HK formula
derived by Holyst and Litniewski. Good agreement was found
for not only monatomic liquids composed of monomers, but also 
molecular liquids composed of dimers and trimers. 
It was confirmed that mechanical equilibrium, i.e., 
a constant component of the stagnation pressure normal to the liquid/vapor interface,
holds for all systems at all times during evaporation.
Furthermore, for liquids evaporating into vacuum, 
the momentum flux contribution dominates in 
the stagnation pressure in the vapor phase,
which was first shown by Holyst and Litniewski to 
lead to the modified HK formula.\cite{holyst09}

Because of evaporative cooling, the temperature of the liquid/vapor interface 
is lower than the bulk temperature of liquid. 
It is further observed that the temperature increases in the vapor phase
with the distance from the liquid/vapor interface. 
This makes the interface the coolest point in the system.
The phenomenon that the vapor phase close to the interface 
has a temperature higher than that 
at the interface is qualitatively consistent 
with a previous experimental measurement on water.\cite{fang99a}

We measured velocity distributions of molecules at the liquid/vapor interface.
As expected, in liquid/vapor equilibrium they follow the MB distribution.
When the evaporation rate is controlled at a small value, 
the distribution also has a MB form but may correspond 
to a temperature that is slightly lower than
the bulk temperature of either liquid or vapor phase. 
When a liquid evaporates into a vacuum, the molecules arriving at the interface
have a velocity distribution that corresponds to a temperature 
much lower than that for the molecules leaving the interface. 
Furthermore, both temperatures 
are lower than the bulk temperature of the liquid phase far from the interface.
In this case, evaporated molecules move away balistically in the vapor phase
without encountering any impedance because the vapor density is extremely low.
It is not surprising that local thermal equilibrium is not reached 
in this extreme situation.
However, our data also showed that under moderate evaporation rates, which
are usually the case in experiments, the vapor density remains 
comparable to its equilibrium value and the hypothesis of local thermal equilibrium 
becomes valid.

The condensation coefficient $\overline{\alpha_c}$ was 
determined by measuring the probability of reflection
of molecules arriving at the liquid/vapor interface from the vapor phase. 
This probability generally depends on the normal component of the velocity 
of the arriving molecules.
The functional form was confirmed to be exponential as in Eq.~(\ref{MBDist}), 
consistent with the model of condensation probability 
of Tsuruta {\it et al.}.\cite{tsuruta95,tsuruta99,tsuruta03,tsuruta05}
Fitting the probability to Eq.~(\ref{MBDist}) 
gave two parameters $\alpha$ and $\beta$, from which
$\overline{\alpha_c}$ was calculated.

The apparent evaporation flux contains two contributions. 
One is from molecules that truly evaporate from the liquid phase, and
another is from molecules arriving at the interface from the vapor phase but being
reflected back. The velocity distributions of these two groups were measured and compared
to modified MB distributions that depend on $\alpha$ and $\beta$. 
From these velocity distributions, $\alpha$, $\beta$ and eventually
the condensation coefficient $\overline{\alpha_c}$ were also determined.

The condensation coefficients $\overline{\alpha_c}$ 
measured with the above two methods are very close, 
which serves as a validation of the kinetic model 
of Tsuruta {\it et al.}.\cite{tsuruta95,tsuruta99,tsuruta03,tsuruta05}
We also found that $\overline{\alpha_c}$ decreases with increasing temperature $T$, 
consistent with the intuition that condensation becomes less likely at higher $T$.
At the same reduced temperature $T/T_c$, 
trimer fluids have the largest $\overline{\alpha_c}$ and monomers have the smallest.
This is in contrast to the traditional viewpoint that monatomic liquids 
have a condensation coefficient close to unity 
and polyatomic liquids have a value less than unity.
However since the chain molecules simulated here are very flexible, 
we do not think our results are in contradiction this viewpoint. 
The chain flexibility makes less prominent 
the rotational degrees of freedom that tend to frustrate the accommodation
of incoming molecules to the liquid phase.
Increasing in chain length leads to a stronger cohesive
interaction between chain molecules. 
As a result, the condensation coefficient is higher for liquids made of molecules 
of longer chains.
However, in most low molecular weight liquids, the molecules are relatively rigid, and the
rotational motion plays a more important role 
in determining liquid cohesion and molecular orientation at the liquid/vapor interface. 
This can reduce the condensation coefficient, 
as suggested by some experiments.
It remains an interesting open question to see if our observation
that larger $\overline{\alpha_c}$ occurs for longer chains can be borne out 
by studying liquids made of a series of chain molecules such as hydrocarbons.

The condensation coefficients $\overline{\alpha_c}$ for monomers, dimers, and trimers
almost collapsed onto a single master curve when plotted against a translational length ratio, 
$(V_l/V_g)^{1/3}$, between the vapor and liquid phase. 
However, all of our data are consistently
above the theoretical prediction of Nagayama and Tsuruta 
based on transition state theory.\cite{tsuruta03}
Note that $V_l/V_g$ is just the inverse of the density ratio.
This deviation indicates that the structure of a liquid/vapor interface
is more complicated than that of a simple material-dividing plane
characterized by a single parameter, $(V_l/V_g)^{1/3}$.

\bigskip\noindent{\bf ACKNOWLEDGMENTS}
\bigskip

This work was made possible by generous allocations of computer time 
at the New Mexico Computing Application Center NMCAC. 
This work is supported by the Laboratory Directed Research and 
Development program at Sandia National Laboratories.
Sandia is a multiprogram laboratory operated by Sandia Corporation, 
a Lockheed Martin Company, for the United States Department of Energy 
under Contract No. DE-AC04-94AL85000.


\end{document}